\documentstyle[twocolumn,aps,epsf]{revtex}

\begin{document}
\draft
\title{Measures of Statistical Complexity: Why?}
\author{David P. Feldman}
\address{Department of Physics, University of California, Davis,
CA 95616\\
Electronic Address: dfeldman@landau.ucdavis.edu}
\author{James P. Crutchfield}
\address{Physics Department,  University of California, Berkeley,
CA 94720-7300\\
and Santa Fe Institute, 1399 Hyde Park Road, Santa Fe, NM 87501\\
Electronic Address:  chaos@gojira.berkeley.edu}
\date{\today}
\maketitle

\bibliographystyle{unsrt}

\begin{abstract}
We review several statistical complexity measures proposed over
the last decade and a half as general indicators of structure
or correlation. Recently, L\`{o}pez-Ruiz, Mancini, and Calbet 
[Phys. Lett. A 209 (1995) 321] introduced another measure of
statistical complexity $C_{{\rm LMC}}$ that, like others, satisfies
the ``boundary conditions'' of vanishing in the extreme ordered and
disordered limits. We examine some properties of $C_{{\rm LMC}}$
and find that it is neither an intensive nor an extensive
thermodynamic variable and that it
vanishes exponentially in the thermodynamic limit for all
one-dimensional finite-range spin systems. We propose a simple
alteration of $C_{{\rm LMC}}$ that renders it extensive. However,
this remedy results in a quantity that is a trivial function of
the entropy density and hence of no use as a measure of structure
or memory. We conclude by suggesting that a useful ``statistical
complexity'' must not only obey the ordered-random boundary
conditions of vanishing, it must also be defined in a setting
that gives a clear interpretation to what structures are quantified.
\end{abstract}

\vspace{-0.2in}


\section{Statistical Complexity Measures}

Theoretical physics has long possessed a general measure of the uncertainty
associated with the behavior of a probabilistic process:  the Shannon
entropy of the underlying distribution \cite{SHANNON,COVER}---a
quantity originally introduced by Boltzmann over $100$ years ago.
In the '50's, Kolmogorov and Sinai \cite{Kolm58,Sina59} adapted Shannon's
information theory to the study of dynamical systems.  This work formed 
the foundation for the statistical characterization of deterministic
sources of apparent randomness in the late '60's through the
early '80's. These efforts to describe the unpredictability of
dynamical systems were largely successful. The metric entropy,
Lyapunov exponents, and fractal dimensions now provide widely
applicable quantities that can be used to detect the presence of
and to quantify the degree of deterministic chaotic behavior.

Since that time, though, it has become better appreciated that
measuring the randomness and unpredictability of a system fails to
adequately capture the correlational structure in its behavior. 
{\em Structure} here is taken to be a statement about the
relationship between a system's components. Roughly speaking, the
larger and more intricate the ``correlations'' between the system's
constituents, the more structured the underlying distribution.
Structure and correlation are not completely independent
of randomness, however.
It is generally agreed that both maximally random and perfectly ordered
systems possess no structure \cite{HUBERMAN,GRASSBERGER,INFERRING}.
Nevertheless, at a given level of randomness away from these extremes,
there can be an enormously wide range of differently structured
processes.  

These realizations led to a considerable effort to develop a general 
measure that quantifies the degree of structure or pattern present 
in a process (c.f. \cite{HUBERMAN,GRASSBERGER,INFERRING,JPC83,SZEPFALUSY,WOLFRAM84,SHAW84,Benn86,LINDGREN88b,LI91,CALCULI,WACKERBAUER,MGM}).
There are many {\em ad hoc} methods for detecting structure,
but none are as widely applicable as entropy is for indicating
randomness.  The
quantities that have been proposed as general structural measures are
often referred to as ``complexity measures.'' To reduce confusion, it
has become convenient to refer to them instead as {\em statistical}
complexity measures. In so doing they are immediately distinguished
from deterministic complexities, such as the Kolmogorov-Chaitin
complexity \cite{Kolm65,Chai66,VITANYI} which requires the deterministic
accounting of every bit---random or not---in an object.  In contrast,
statistical complexity measures discount for randomness and so provide
a measure of the regularities present in an object above and beyond pure
randomness. Deterministic complexities are dominated by the random
components in an object; the result is that their average-case growth
rate is given by the Shannon entropy rate \cite{COVER}.

A number of approaches to measuring statistical complexity have 
been taken. One line of attack operates 
within information theory and examines how the Shannon entropy of
successively larger subsystems converges to the entropy density of the
entire system \cite{GRASSBERGER,JPC83,SZEPFALUSY,SHAW84,LINDGREN88b,LI91}.
These quantities can be interpreted as the average
memory stored in configurations.

Another set of approaches appeals to computation theory's classification
of devices that recognize different classes of formal language (a set of
strings).  Examples include finite memory devices (e.g. the finite-state
machines) and infinite memory devices (e.g. push-down automata and Turing
machines) \cite{Hopc79}. One such computation-based measure of statistical
complexity is the {\em logical depth} \cite{Benn86}. The logical depth of
(say) a system's configuration is the time required for a universal Turing
machine to run the minimal program that reproduces it. Another example of
a computation theoretic approach is found in the statistical complexity of
refs.~\cite{INFERRING,CALCULI}, a quantity that measures the amount of
memory needed, on average, to statistically reproduce a given configuration.
Unlike logical depth, which assumes the use of a universal Turing
machine---the most powerful discrete computational model class---this 
statistical complexity assumes that the simplest possible computational 
class is used to describe the configuration. A higher level class is used 
only if lower ones fail to admit a finite description. This ``bottom-up'' 
definition of hierarchical information processing has been successfully 
applied to the symbolic dynamics of chaotic maps \cite{INFERRING,CALCULI}, 
cellular automata \cite{TURBULENT}, spin systems \cite{JPC.DPF.1}, and 
hidden Markov models \cite{UPPER-THESIS}. For other approaches to 
statistical complexity see, for example,
refs.~\cite{HUBERMAN,WOLFRAM84,WACKERBAUER,MGM,Lloy88}.

\section{Properties of $C_{\rm LMC}$}

Recently, L\`{o}pez-Ruiz, Mancini, and Calbet proposed another
measure of statistical complexity $C_{{\rm LMC}}$ \cite{LMC}.
Consider a discrete random variable $Y$ that can take on $N$
values $y$.  We denote by ${\rm Pr}(y)$ the probability that
the variable $Y$ assumes the value $y$. Ref.~\cite{LMC} then
defines a complexity measure:
\begin{equation}
        C_{{\rm LMC}}[Y] \, \equiv \, H[Y]\,D[Y] \;,
\label{C.def}
\end{equation}
where $H$ is the Shannon entropy,
\begin{equation}
        H[Y]\, = \, -\sum_{\{y\}}{\rm Pr}(y) \log_2 {\rm Pr}(y) \; ,
\end{equation}
in which the sum runs over all allowed values of $y$.
The quantity $D$ is the ``disequilibrium,'' defined by
\begin{equation}
        D[Y] \, = \, \sum_{\{y\}}  ( {\rm Pr}(y) - \frac{1}{N} )^2  \;,
\label{D.def}
\end{equation}
which measures the departure of ${\rm Pr}(y)$ from uniformity.

The motivation posited in ref. \cite{LMC} for the form of
$C_{{\rm LMC}}$ is that it vanishes for distributions that correspond
to perfect order and maximal randomness.
Ref.~\cite{LMC} argues that perfect order corresponds to a vanishing
Shannon entropy and notes that for $H = 0$, $C_{{\rm LMC}} = 0$.
Maximal randomness occurs for $ H = \log_2 N$, corresponding to
${\rm Pr}(y) = 1/N$. And so, by eq.~(\ref{D.def}) $D$ and hence
$C_{{\rm LMC}}$ equal zero. Thus, by construction, $C_{\rm LMC}$
vanishes in the extreme ordered and disordered limits.

We now proceed to discuss the behavior of $C_{\rm LMC}$ in the
thermodynamic limit.  Anteneodo and Plastino have already reported
some of $C_{\rm LMC}$'s properties in this limit \cite{PLASTINO}.  
However, their line of investigation is rather different from that
undertaken here. In ref.~\cite{PLASTINO} the distribution that
maximizes $C_{\rm LMC}$ is determined. Numerical and analytic work
there indicate that the maximizing distribution for $N \rightarrow \infty$
is one in which a single event has probability $2/3$ while all others
are equally likely.

As an alternative to looking at the maximizing distribution, we suggest
examining how a system's complexity changes as parameters---e.g
temperature, coupling strength, nonlinearity, etc.---are varied.  
This approach is in keeping with statistical mechanics, where
one typically looks at changes in the behavior of 
quantities in the thermodynamic limit as system parameters are varied.
For example, rather than determine the distribution that maximizes the
expectation value of the specific heat for a finite-sized system, one
usually determines how the specific heat per site behaves in the limit
of an infinite system as, say, the temperature is varied.

Consider the class of probability distributions over discrete, finite
random variables generated by finite-memory Markov chains. Let
$\ldots X_{-2}, X_{-1}, X_0, X_1, X_2, \ldots $ be a bi-infinite
chain of random variables where each value $x_i$ is chosen from a
discrete finite alphabet of size $k$. We denote $L$ consecutive
variables by $X_i^L \equiv X_i, X_{i+1}, X_{i+2}, \ldots, X_{i+L-1}$.
$X_i^L$ is a system of $L$ variables with $N = k^L$ possible
configurations $x_i^L$.  Let ${\rm Pr}(x_i)$ be the probability that
the $i^{\rm th}$ random variable takes on the particular value $x_i$.
We denote by ${\rm Pr}(x_i^L)$ the joint probability distribution over
$L$ consecutive random variables. We assume a shift symmetry so that
${\rm Pr}(x_i^L) = {\rm Pr}(x_0^L)$ and subsequently drop the subscript
$i = 0$. The chain of discrete random variables may be viewed as a
translationally invariant spin system or, equivalently, as a stationary
stochastic process.

As is well known, the Shannon entropy of a block of $L$ such variables
typically grows linearly for sufficiently large $L$.  In other words, 
the limit
\begin{equation}
    h_\mu \, \equiv \, 
        \lim_{L \rightarrow \infty} \frac{1}{L}
        H[X_0, X_{1}, \ldots , X_{L-1}]  
\label{hmu.def}
\end{equation}
exists and, in the thermodynamic ($L \rightarrow \infty$) limit,
\begin{equation}
    H[X^L] \, \propto \, h_\mu L \; .
\end{equation}
The quantity $h_\mu$ is well-defined as the system size goes to
infinity and is known as the entropy rate, the metric entropy, or
the thermodynamic entropy density depending on the context. In
statistical mechanics parlance, eq.~(\ref{hmu.def}) tells us that
the Shannon entropy $H$ is an extensive quantity, so that
it is possible to define a meaningful entropy density $h_\mu$ that
characterizes the randomness per variable in the system.

The ``disequilibrium'' term, eq.~(\ref{D.def}), is not so well behaved.
In fact, under no circumstances does it grow linearly with system size $L$.  
One can show, quite generally, that for a system of length $L$ described by
a probability distribution over $N = k^L$ events, $D$ is bounded above by
$1 - 1/N$.  To see this, we expand the square in eq.~(\ref{D.def}) and,
since the probability distribution is normalized, 
obtain
\begin{equation}
    D[X^L] \, = \, \sum_{ \{x^L \}} {\rm Pr}(x^L)^2 - 
        \frac{1}{k^L} \;.
\label{D.expanded}
\end{equation}
The sum is understood to run over all $k^L$ possible values of $X^L$.
Since ${\rm Pr}(x^L) \leq 1$, it follows that 
\begin{equation}
    \sum_{ \{x^L \}} {\rm Pr}(x^L)^2 \, \leq \,  \sum_{ \{x^L \}}
        {\rm Pr}(x^L) \, = \, 1 \;.
\end{equation}
Thus, 
\begin{equation}
    D[X^L] \, \leq \, 1 - k^{-L} \;, 
\label{D.bound}
\end{equation}
and we see that $D$ cannot grow linearly with the system size $L$. 
Therefore, the ``disequilibrium'' is not a thermodynamically 
extensive quantity.

In fact, it can be shown that $D$ vanishes exponentially with increasing
system size for a large class of systems.  Let our chain of variables
$\ldots X_{-1}, X_0, X_1, X_2, \ldots$ be chosen by a one-step Markov
process with transition probabilities given by
$T_{ab} = {\rm Pr}(b|a) , ~a,b \in \lbrace 0, 1, \ldots , k-1 \rbrace$.
That is, $T_{ab}$ gives the probability that the variable $X_{i+1}$
takes on the $b^{\underline {\rm th}}$ value given that $X_i$ takes
on the $a^{\underline {\rm th}}$ value.
(What follows applies to any finite-step Markov chain, as blocks of 
adjacent variables can be grouped to render the process one-step.)
Then, if we assume that the Markovian process is regular---{\em i.e.},
there exists some $K$ such that $(T_{ab})^K > 0$ for all $a$ and
$b$---then it follows that the disequilibrium of a system of $L$
variables goes to zero exponentially fast in $L$. This is proved
in appendix \ref{exponential.vanishing}.  

As a result, $C_{\rm LMC}$ vanishes in the thermodynamic limit for 
all regular Markov chains, a class of systems that includes
all finite-range, one-dimensional spin systems with
finite-strength interactions. It seems to us counterintuitive that
a (useful) measure of complexity vanishes for all of these systems.
While these models exhibit no critical phenomena, there are considerable
changes in the structure of the distributions as system parameters
are varied \cite{JPC.DPF.1}. A measure of complexity, as we envision
it, should be sensitive to these changes.

As an illustration of the nonextensivity of $D$, consider the special
case where the chain consists of variables that are independent
and identically distributed (iid); {\em i.e.},  
${\rm Pr}(x_i, x_j) \, = \, {\rm Pr}(x_i) {\rm Pr}(x_j)$ for
all $X_i$ and $X_j$ with $i, j = \ldots , -2, -1, 0, 1, 2, \ldots $.
For a spin system, this corresponds to the case where there is no
coupling between spins---a paramagnet.  For convenience we assume 
that $X_i$ can take on two values, (say) $0$ and $1$, and we denote
the probability that $X_i$ takes on the value $1$ by $p$. Then, using
the binomial theorem, we find that $C_{LMC}$ for a system of
$L$ such variables is given by:
\begin{equation}
    C_{{\rm LMC}}[X^L] \, = \, L H[X] \, \left(  
        ( 1 - 2p + 2p^2)^L \, - \, 2^{-L} \right) \; ,
\label{C.binomial}
\end{equation}
where $H[X]$ is the binary entropy function:
\begin{equation}
    H[X] = - p \log_2 p \, - \, (1-p)\log_2 (1-p) \;.
\label{H.binomial}
\end{equation}

Eq. (\ref{C.binomial}) is rather curious.
In our view, the complexity of a collection of iid binary variables 
should vanish regardless of their number.
A set of independent variables is statistically very simple---there
is manifestly no correlational structure whatsoever.
Furthermore, it seems to us that if the complexity doesn't vanish
for all such systems, it ought to grow linearly as a function of
the number of variables.  That is, six biased coins should be twice as 
complex as three biased coins.  Eq.~(\ref{C.binomial}) shows that
the size dependence of $C_{\rm LMC}$ is much more complicated.

In ref.~\cite{PLASTINO} it is found that the maximal value of 
$C_{{\rm LMC}}$ goes to $4/27$ as the system size goes to infinity.
This is not at odds with the exponentially fast vanishing of $D$
(and hence $C_{{\rm LMC}}$) for regular Markov processes
noted here, since the system that maximizes $C_{\rm LMC}$
is not Markovian. To see this, recall that the maximal distribution
reported in ref.~\cite{PLASTINO} has one configuration with probability
$2/3$ while all others have equal probability. In the thermodynamic
limit, this one configuration is infinitely long. Thus, the generating
process must keep track of arbitrarily long sequences in order to
assign a distinct probability to one and another probability
uniformly to all others.
As a result, the distribution that maximizes $C_{\rm LMC}$
cannot be generated by a finite-memory Markov process in the
thermodynamic limit.

\section{Repairing Nonextensivity}
Up to this point we have seen that $C_{\rm LMC}$ is not
suitable for use in a statistical mechanics context. In particular,
it suffers from two related deficiencies: it is not an extensive
quantity and it vanishes for a large variety of structured processes.
The trouble causing both of these shortcomings resides in the 
``disequilibrium'' factor. Perhaps if one altered the definition 
of $C_{\rm LMC}$ so that $D$ was extensive, as the Shannon entropy
$H$ is, one would obtain a more useful measure of statistical
complexity.

To this end, we seek an extensive measure of a distribution's departure
from uniformity. As we will be multiplying this measure by the Shannon
entropy $H$ which carries units of bits, it also seems natural,
although not necessary, to choose a ``disequilibrium-like''
quantity that also carries units of bits. Information
theory is armed with just such a function: the relative information
\cite{COVER}.

The relative information, also known as the information gain or the
Kullback-Leibler information distance, between two distributions
${\rm Pr}(y)$ and $\widehat{\rm Pr} (y)$ is defined by
\begin{equation}
    D( \, {\rm Pr}(y)\, \| \, \widehat{\rm Pr} (y)\, ) \, \equiv \, 
        \sum_{ \{ y\} } {\rm Pr}(y) 
        \log_2 \frac{{\rm Pr}(y)}{\widehat{\rm Pr} (y)}\; .
\label{rel.ent.def}
\end{equation}
The relative information is not a true distance function---neither
satisfying the triangle inequality nor being symmetric. Nevertheless,
it does provide a measure of how much two distributions differ
and it does carry the same units (bits) as the Shannon entropy.
It is also an extensive quantity, since it grows linearly with
the number of variables in the distribution's support.

So, $D( \, {\rm Pr}(y)\, \| \, \widehat{\rm Pr} (y)\, )$ where 
$\widehat{\rm Pr} (y) = 1/N $ provides an extensive measure of
${\rm Pr}(y)$'s departure from uniformity in units of bits.
Using this in eq.~(\ref{D.def}), we define a modified statistical
complexity measure:
\begin{equation}
     C^{{\, \prime}}[Y] \,\equiv 
        \, H[Y] \, D( \, {\rm Pr}(y) \, \| \, 1/N \, )  \; ,
\label{modified.C.def}
\end{equation} 
which has units of $[{\rm bits} ^2 ]$.
From here on we will focus on $C^{\prime}$, the modified $C_{\rm LMC}$.
Note that much of $C^{\prime}$'s character is shared by $C_{\rm LMC}$.

To see how this new quantity behaves, let's look more closely at
$D(\, {\rm Pr}(y) \, \| \, 1/N \,)$.  Consider again $X^L$, a Markov
chain of length $L$. And for convenience let the $X_i$ be binary
variables. The total number $N$ of configurations for such a system
is $2^L$. First, note that:
\begin{equation}
   D( \, {\rm Pr}(x^L) \, \| \, 1/N \, ) \, = \, L - H[ X^L ] \; .
\label{rel.ent.uniform}
\end{equation}
Since $H[X^L]$ is extensive, we have:
\begin{equation}
   D( \, {\rm Pr}(x^L) \, \| \, 1/N \, ) \, \propto \,
        L (1 - h_\mu) \; .
\end{equation}

\begin{figure}[tbp]
\epsfxsize=3.0in
\epsfysize=3.0in
\epsffile{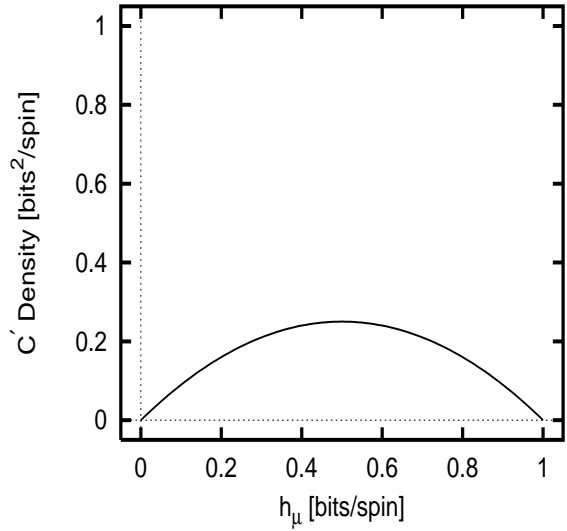}
\caption{$C^{\, \prime}$ versus entropy density $h_\mu$.  Note that
$C^{\, \prime}$ is a function of $h_\mu$.
}
\label{CPrime.vs.hmu}
\end{figure}

Thus, $C^{\, \prime}$ consists of the product of two extensive
quantities, $H$ and $D( \, {\rm Pr}(x^L) \, \| \, 1/N \, )$. As
a consequence, dividing $C^{\, \prime}$ by $L^2$ yields a quantity
that is finite in the $L \rightarrow \infty$ limit; specifically,
\begin{equation}
   \lim_{L \rightarrow \infty} \frac{1}{L^2} C^{\, \prime} \, = \,
        h_\mu (1 - h_\mu) \; .
\label{C.prime.is.entropy}
\end{equation}
(See Fig. \ref{CPrime.vs.hmu}.)
This is indeed a function that vanishes in the ordered ($h_\mu = 0$) and
disordered ($h_\mu = 1$) extremes.  However, note that it is a function
{\em only}\/ of the system's entropy density.  That is, the modified
statistical complexity measure eq.~(\ref{modified.C.def}) is a function
only of the system's randomness. 
The relation  $C^{\, \prime} \propto  L^2 h_\mu ( 1 - h_\mu)$ strikes
us as being ``over-universal.''
For example, it's possible for a paramagnet and a system at its critical
point, where the correlation length diverges, to have the same value of
$C^{\, \prime}$.  It does not seem particularly revealing
that all systems with the same entropy density have the same statistical
complexity.  

\begin{figure}[tbp]
\epsfxsize=3.0in
\epsfysize=3.0in
\epsffile{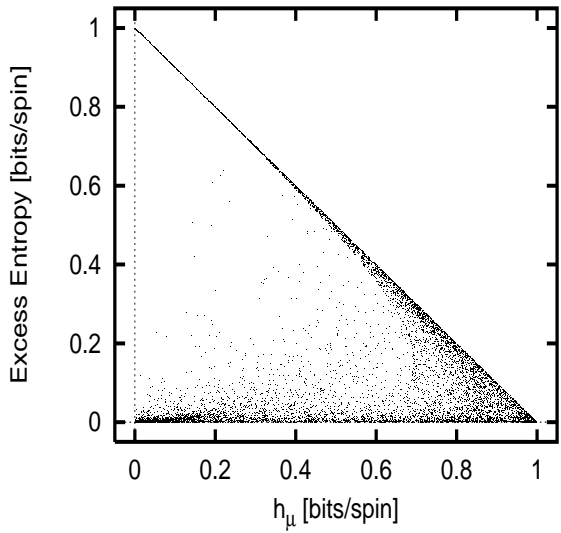}
\caption{Excess entropy ${\rm E}$, a statistical complexity measure,
versus entropy density $h_\mu$ for a 
spin-$1/2$ one-dimensional Ising spin system: $10^4$ ($h_\mu$, $\rm E$)
points. The system parameters were randomly chosen from the following
intervals:  $J$ (coupling constant) $\in [-3,3]$; $T$ (temperature)
$\in [0.05,4.05]$; and $B$ (external field) $\in [0,3]$.
}
\label{E.vs.hmu}
\end{figure}

As a contrast to the $C^{\, \prime}$ versus $h_\mu$ behavior shown in
Fig.~(\ref{CPrime.vs.hmu}), consider Fig.~(\ref{E.vs.hmu}), where
we show the behavior of a different measure of statistical complexity,
the {\em excess entropy} $ \rm E$
\cite{GRASSBERGER,JPC83,SZEPFALUSY,SHAW84,LINDGREN88b,LI91}.
The excess entropy of an infinite configuration may be expressed as
the mutual information between two semi-infinite halves of the
configuration. That is, $\rm E$ is the amount of spatial memory embedded
in configurations.  In ref.~\cite{JPC.DPF.1} we show that ${\rm E}$
captures significant structural changes in the configurations of
one-dimensional spin systems as external parameters are varied. We have
plotted the excess entropy for $10^4$ sets of parameter values for a 1D
Ising system with nearest-neighbor coupling in the presence of an
external field. Note that for {\em all} the Ising systems plotted in
Fig.~(\ref{E.vs.hmu}), $C_{\rm LMC} = 0$, since $C_{\rm LMC}$ vanishes in
the thermodynamic limit. Comparing the two figures, it is clear that the
excess entropy depends on the entropy density $h_\mu$ in a much more
subtle way than $C^{\, \prime}$ does.  

Comparing these two plots raises another important issue. Note that
the excess entropy does not always equal zero for $h_\mu = 0$,
an apparent violation of the ``boundary condition'' requiring that
a complexity measure vanish in the perfectly ordered limit. However,
$h_\mu = 0$ corresponds to perfect {\em asymptotic predictability},
not perfect {\em order.} A process with a vanishing entropy rate indicates
that it can be predicted without error---it says nothing, however, about
how much effort or memory is required to perform this prediction. Thus, a
zero value of the entropy density is too crude a measure of order.
To see this, note that {\em any} periodic system has $h_\mu = 0$.
Yet all periodic systems aren't equally ordered: a configuration
with period $1$ is certainly more ordered than a configuration of
period $1729$---which, for example, requires more memory to produce.
In fact, at the period-doubling accumulation point of the logistic
map, the symbolic dynamics produce periodic configurations of
diverging periodicity. Hence, the excess entropy is infinite here,
while the entropy rate remains zero
\cite{GRASSBERGER,INFERRING,CALCULI}.

In contrast to $C^{\, \prime}$, which vanishes for any periodic
system, the excess entropy $\rm E$ for a configuration of period
$P$ is $\log_2 P$.  Only if the period is $1$, indicating trivial
ordering and predictability, does the excess entropy vanish for a
periodic process. Thus, the $(h_{\mu},{\rm E}) = (0,0)$ points
in Fig.~(\ref{E.vs.hmu}) correspond to the system's ferromagnetic
ground states of period 1 and the $(h_{\mu},{\rm E}) = (0,1)$ points
correspond to the antiferromagnetic ground states of period 2
\cite{JPC.DPF.1}. Statistical complexity measures such as
$C^{\, \prime}$ or, for that matter $C_{\rm LMC}$, that are zero for
all $h_\mu = 0$ configurations are very blunt implements with which
to detect structure. A measure of complexity should be able to
distinguish between structures of different periodicities. 

For maximal randomness ($h_\mu = 1$), the excess entropy ${\rm E}$
vanishes, as expected.  At $h_\mu = 1$, corresponding to infinite
temperature, the spins decouple and there is no information shared 
between them.  But the excess entropy does more than satisfy the
``boundary conditions'' of vanishing for $h_\mu = 0$ and $1$.  The
interpretation of ${\bf E}$ as the memory stored in spatial configurations
holds for intermediate values of $h_\mu$ as well.  As a result, 
Fig.~(\ref{E.vs.hmu}) lets us place an upper bound on the memory stored
in spatial configurations for a spin-1/2 nearest neighbor Ising model:
$ {\rm E} \leq 1 - h_\mu$.  This result, derived analytically in
ref.~\cite{JPC.DPF.1}, applies to all one-step Markov chains over a
binary alphabet.

We conclude this section by noting that there is a growing body of
evidence indicating that, aside from the requirement of vanishing at
the ordered and disordered extremes, entropy and ``complexity''
(defined in a number of different ways to reflect ``structure'')
are more or less independent. That is, there is a vastly wider range
of complexity versus entropy relationships than indicated by
eq.~(\ref{C.prime.is.entropy}) \cite{LI91,CALCULI,JPC.DPF.1}.
Fig.~(\ref{E.vs.hmu}) is just one example of many possible statistical
complexity-entropy density relationships.

\section{Conclusion}

To summarize, we have shown that $C_{\rm LMC}$ vanishes in the
thermodynamic limit for finite-memory regular Markov chains.
This class of systems includes, at a minimum,
all finite-range one-dimensional
spin systems.  We have also shown that $C_{\rm LMC}$ is not
an extensive variable. 
We have proposed modifying $C_{\rm LMC}$, replacing the
``disequilibrium'' of ref.~\cite{LMC} with the relative entropy with
respect to the uniform distribution.  This results in an
quantity $C^{\, \prime}$ that grows appropriately in the thermodynamic
limit, making it possible to define a meaningful statistical complexity
density that, nonetheless, retains the spirit of $C_{\rm LMC}$.
However, the product of this modification is a quantity that
is a trivial, ``over-universal'' function of the entropy density
$h_\mu$.  In short, based on the above observations, it seems to us 
that $C_{\rm LMC}$ and $C^{\, \prime}$ may be of little use in measuring
the complexity of a statistical mechanical system.

We conclude by pointing out that the ``boundary
conditions'' of vanishing in the extreme ordered and disordered limits
do not uniquely specify a measure of complexity, an observation also
made by Anteneodo and Plastino \cite{PLASTINO}.
In fact, if this is the
only feature one demands of a complexity measure, it's not clear
to us why one would be motivated to devise a new statistic at all.

Statistical mechanics, for example, is replete with functions that
vanish in the high and low temperature limits. Since thermodynamic
entropy, a measure of randomness, is a monotonic function of
temperature, high (low) temperature corresponds to high (low)
randomness. Examples of quantities that vanish in these extremes
(assuming there is not a critical point at $T = 0$) include the
connected correlation functions, the correlation length, and
magnetic susceptibility. These functions can be easily applied
to any probability distribution describing a spatially or
spatio-temporally extended collection of random variables.

Information theory also comes equipped with a function that vanishes
for perfectly ordered and disordered systems:  the mutual information
$I$ \cite{SHANNON,COVER}. If two random variables $Z$ and $Y$ are
independently distributed, then the mutual information between them,
$I[Z;Y]$ vanishes. At the other extreme, if $Z$ and $Y$ are both known
with certainty---that is, $H[Z] = H[Y] = 0$---then $I[Z;Y]$ also
vanishes. For statistical dependencies between these extremes,
$I[Z;Y]$ is positive and measures the amount of information shared
between $Y$ and $Z$. 

Given that there are many functions that vanish in the extreme ordered
and disordered limits, it is clear that requiring this property
does not sufficiently restrict a statistical complexity measure.  
What other criteria can we use, then, to guide us as we attempt to
detect structures and patterns in nature? To this question we
offer two suggestions.

First, it is helpful if the statistical complexity measure has a clear
interpretation: {\em What exactly is the statistical complexity
measuring?} The two English words ``statistical complexity'' do not
sufficiently answer the question.  Many of the statistical complexity
measures proposed over the last decade or so do have clear
interpretations. For example, the excess entropy may be interpreted
as the mutual information between two halves of an infinite
configuration \cite{GRASSBERGER,LI91,JPC.DPF.1}. Logical depth is the
run time required by a universal Turing machine executing the minimal
program to reproduce a given pattern. These unambiguous interpretations
help put these statistical complexity measures on a solid footing.

Second, it is essential to consider the motivations behind a measure
of statistical complexity: {\em How is the measure to be used? What
questions might it help answer?} It is possible to meaningfully assess
its utility only if the motivations and goals for defining a
complexity measure are stated clearly. One set
of issues is the detection and quantification of patterns produced
by a process. It has been proposed that particular notions of
structure adapted from computation theory capture the intrinsic
``patterns'' and information processing architecture embedded in a
system. In this setting, one finds well-defined and easily
interpreted measures of statistical complexity \cite{CALCULI}.
For other views on questions that a measure of statistical complexity
might help answer, see ref.~\cite{Benn90}.

Finally, ref.~\cite{PLASTINO} mentions several different notions of
complexity and notes that ``there is not yet a consensus on a precise
definition.'' ``Complexity'' has accepted meanings in other
fields---meanings established prior to the recent attempts to use
it as a label for structure in natural systems. For example,
{\em Kolmogorov-Chaitin complexity} in algorithmic information theory
means something quite different from {\em computational complexity}
in the analysis of algorithms. These, in turn, are each different from
the {\em stochastic complexity} used in model-order estimation in
statistics \cite{RISSANEN}. Though at a future date relationships may
be found, at present all of these are different from the notions of
statistical complexity discussed here.

Unfortunately, ``complexity'' has been used without qualification
by so many authors, both scientific and non-scientific, that
the term has been almost stripped of its meaning. Given this state
of affairs, it is even more imperative to state clearly why one is
defining a measure of complexity and what it is intended to capture.

We thank Melanie Mitchell and Karl Young for helpful comments on the
manuscript. This work was partially supported at UC Berkeley by ONR
grant N00014-95-1-0524 and at the Santa Fe Institute by ONR grant
N00014-95-1-0975.

\vspace{-0.2in}
\bibliography{mscw}
\vspace{-0.2in}

\appendix
\section{$D$ vanishes exponentially fast for
Regular Markov Chains}
\label{exponential.vanishing}
We will show that $D$ goes to zero exponentially fast with increasing
system size for a Markov chain governed by a regular transition matrix
$T_{ab} \, = \, {\rm Pr}(b|a)$, where $0 \leq T_{ab} \leq 1$ and there
exists a $K < \infty$ such that $(T^K)_{ab} > 0$ for all $a$ and $b$.
Since the conditional probabilities are normalized, the matrix $T$ is
stochastic: $\sum_{b=1}^k T_{ab} = 1$.

The probability of a block of $L$ consecutive variables taking on
the values $x_1, x_2, \ldots, x_L$ is given by
\begin{equation}
    {\rm Pr}(x_1, x_{2}, \ldots , x_{L}) \, = \, p_{x_1} T_{x_1 x_2}
        T_{x_2 x_3} \cdots T_{x_{L-1}x_L} \; ,
\end{equation}
where $p$ is the stationary distribution of a single variable, as given
by the left eigenvector of $T$ with eigenvalue $1$. The eigenvector $p$
is chosen so as to be normalized in probability, $\sum_{a=1}^k p_a = 1$.

Let $\tilde{T}$ be a matrix whose components $\tilde{T}_{ab}$ are given 
by $(T_{ab})^2$. Note that $\tilde{T} \neq T^2$. Similarly, let
$\tilde{p}$ be a vector whose components $\tilde{p}_a$ are given by
$(p_a)^2$. Eq.~(\ref{D.expanded}) indicates that we are interested in
\begin{equation}
\sum_{ \{ x^L \} } {\rm Pr}(x^L)^2
  = \sum_{ \{ x^L \} }
  \tilde{p}_{x_1} \tilde{T}_{x_1 x_2} \tilde{T}_{x_2 x_3} 
  \cdots \tilde{T}_{x_{L-1}x_L} .
\end{equation}
The sum runs over all configurations of length $L$.  
The effect of the sum is to multiply the matrices together;
\begin{equation}
    \sum_{ \{ x^L \} } {\rm Pr}(x^L)^2 \, = \, \sum_{x_1} \sum_{x_L}
        \tilde{p}_{x_1}(\tilde{T}^{L-1})_{x_1 x_L} \; .
\label{matrix.multiply}
\end{equation}
We shall show that in the $L \rightarrow \infty$ limit the above
expression goes to zero exponentially fast. 

We begin by considering the vector $V \equiv \tilde{p} \tilde{T}^{L-1}$ and
its $L_\infty$ norm, $\| V \| \equiv \max \{ |V_1|, |V_2|, \ldots \}$.
Eq.~(\ref{matrix.multiply}) may be rewritten in terms of $V$,
\begin{equation}
    \sum_{ \{ x^L \} } {\rm Pr}(x^L)^2 \, = \, \sum_{i=1}^k V_i .
\end{equation}
Since $V$ is finite dimensional and all elements are nonnegative,
if $ \| V \| $ goes to zero exponentially fast,
$\sum_{ \{ x^L \} } {\rm Pr}(x^L)^2 $ must also go to zero
exponentially fast. To show the former we use some well-known
properties of vector and matrix norms \cite{BRONSON}.

Consider the matrix norm induced by the $L_\infty$ vector norm:
\begin{equation}
    \| \tilde{T} \| \, \equiv \,
        \max_a \{ \, \sum_b \tilde{T}_{ab}\, \} \;.
\end{equation}
Any matrix norm is compatible with its associated vector norm:
\begin{equation}
        \|V\| \, = \, \| \tilde{p} \, \tilde{T}^{L-1} \| \,
        \leq \, \| \tilde{p} \| \, \| \tilde{T}^{L-1} \| \;.
\end{equation}
Recall that the components of $\tilde{p}$ are the square of the components
of the stationary probability $p$ of the Markov chain. 
Except for the trivial case in which there is only one symbol in our
chain and $T$ is a one-by-one matrix, the maximum component 
of $p$ is less than one. Thus, $0< \| \tilde{p} \|< 1$ and we have
\begin{equation}
    \| V \| \, \leq \,  \| \tilde{p} \| \, 
        \| \tilde{T}^{L-1} \| \, \leq \, \| \tilde{T}^{L-1} \| \;.
\label{inequality}
\end{equation}

By assumption, there exists
a $K$ such that $0 < (T^K)_{ab} < 1 $ for all $a$ and $b$. Each
element of $T^K$ is a sum of terms that are products of $T$'s elements.
Likewise, each element of $\tilde{T}^K$ is a sum of terms that are
products of $\tilde{T}$'s elements. However, since
$\tilde{T}_{ab} = (T^2)_{ab}$ and $ 0 \leq T_{ab} \leq 1$, it follows
that each component of $\tilde{T}^K$ is strictly less than the
corresponding component of $T^K$. The product of stochastic matrices
is itself a stochastic matrix, so $\sum_{b=1}^k (T^K)_{ab} = 1$. Thus,
since each component of $\tilde{T}$ is less than the corresponding
component of $T$, $\sum_{b=1}^k (\tilde{T}^K)_{ab} < 1$. As a result,
$\| \tilde{T} \| < 1$.

Now, by the consistency condition obeyed by all matrix norms,
$\| \tilde{T}^2 \| \leq \|\tilde{T} \| \, \| \tilde{T} \| $.
Rewriting eq.~(\ref{inequality}), we have:
\begin{equation}
        \| V \| \leq \| \tilde{T}^{L-1} \| =
        \| \tilde{T}^{K(L-1)/K} \| \; .
\end{equation}
In the $L/K \rightarrow \infty$ limit---equivalent to the $L \rightarrow
\infty$ limit since $K$ is finite---we then have by the consistency
condition:
\begin{equation}
        \| V \| \leq \| \tilde{T}^K \|^{L-1} \; .
\end{equation} 
Since  $ \| \tilde{T}^K \| < 1$ we see that $\|V \|$ is
bounded above by a function that decreases exponentially in $L$. Hence
$\| V \|$ itself also decreases exponentially in $L$.

\end{document}